# Conductivity-driven chirality in geometrically achiral plasmonic nanostructures and metasurfaces

Yulong Duan[1§*], Longzhu Liu[2§], Suraya Kazi[1], Harry Miyosi Silalahi[1], Roger Magnusson[3], Xing Xing[1], Zhe Zhang[4], Dongqing Lin[1], Wanlin Wang[4] and Magnus P. Jonsson[1*]

[1]Laboratory of Organic Electronics, Department of Science and Technology (ITN), Linköping University, Norrköping, SE-60174, Sweden

*E-mail: yulong.duan@liu.se, magnus.jonsson@liu.se

[2]Faculty of Arts and Sciences, Beijing Normal University, Zhuhai 519087, China

[3]Department of Physics, Chemistry and Biology (IFM), Linköping University, SE-58183, Sweden

[4]School of Physical Sciences, Great Bay University, Dongguan, 523000, China

[§]These authors contributed equally

## Abstract

**Resonant nanostructures are powerful for chiral control of light, but widely assumed to require achiral geometry. Here we demonstrate that optical chirality can emerge in geometrically achiral nanostructures, through the misalignment of independent linear responses. By rotating the conductivity tensor of an anisotropic plasmonic medium relative to a geometrically symmetric nanostructure, we generate a spin-selective optical response that can be continuously tuned both in magnitude and handedness, including reversible switching between left- and right-handed behaviour. We experimentally realize this concept using an aligned plasmonic conducting polymer and by controlling the orientation of its polymer backbone relative to otherwise identical rectangular nanoantennas. This leads to strong planar chirality that can also be reversibly modulated by varying the carrier density of the polymer. Moreover, the rotation angle of rectangular meta-atoms made through this principle encodes no longer only the local geometric phase but simultaneously the local optical chirality. This enables a new design paradigm for phase-gradient metasurfaces , enabling simultaneous control of optical chirality and wavefronts through a single rotational degree of freedom.**

## Introduction

Chiral light–matter interactions play a central role in molecular spectroscopy, polarization control and various emerging photonic applications[1]. In turn, achieving chiroptical control needs broken mirror symmetry, which can be provided by geometrically chiral three-dimensional nanostructures[2] or twist-stacked arrangements of linearly anisotropic layers[3-5]. In molecular systems, mirror symmetry-breaking can originate from inherent chirality of the constituent molecules, supramolecular packing into chiral arrangements, or anisotropic arrangements that induce misalignment between the molecular transition dipole and the birefringency axis[7-8]. Notably, the chiral optical response of such molecular films generally remains weak because appreciable phase retardation between left-handed and right-handed circularly polarized light (LCP and RCP, respectively) requires large film thicknesses. By contrast, planar metasurfaces composed of resonant nanoantennas can exhibit strong circular dichroism (CD) at deep subwavelength thicknesses. However, such systems have relied on nanoantennas that are geometrically chiral,[9-10] composed of multiple materials[11], or arranged into chiral lattices[12-13], or have required extrinsic chiral configurations in which oblique incidence illumination geometry breaks the symmetry of the system[14]. Hence, the combination of strong chiroptical response, normal-incidence operation, geometric symmetry, and a single-material implementation within a planar architecture remains an unmet goal.

Here, we introduce a general principle for realizing optical chirality in geometrically achiral resonant nanostructures under normal-incidence illumination. Starting from a geometrically achiral planar plasmonic structure, such as a rectangular nanobar, symmetry is broken not be geometry but by anisotropic material conductivity, here provided by an aligned  conducting polymer. The key is to misalign the principal axis of the conductivity tensor relative to the geometric symmetry axis of the nanostructure. This misalignment breaks mirror symmetry and induces chiroptical response that can be controlled in both magnitude and sign via the misalignment angle (Fig. 1). We show theoretically and experimentally that this general principle applies to different achiral geometries, including regular linear gratings. When based on an anisotropic conducting polymer, the effect can also be reversible switched off and on by modulating the charge carrier density of the polymer.

We further show that our mechanism fundamentally changes the physical meaning of the rotation angle in metasurfaces based on the Pancharatnam–Berry (PB) phase.[15] Unlike conventional PB metasurfaces, where the rotation angle serves solely to encode the geometric phase, the rotation of an anisotropically conductive rectangular meta-atom relative to its principal conductivity axis simultaneously determines both the geometric phase and the local optical chirality. As a result, the prescribed phase profile is intrinsically accompanied by a deterministic spatial chirality texture, which is absent in conventional PB metasurfaces based on isotropic materials. This previously unexplored coupling between geometric phase and chirality introduces an additional degree of freedom for metasurface design, enabling full phase control while simultaneously introducing chiral asymmetry present in the farfield. We exemplify this capability *via* metagratings that provide beam deflection with controlled chiral asymmetry.

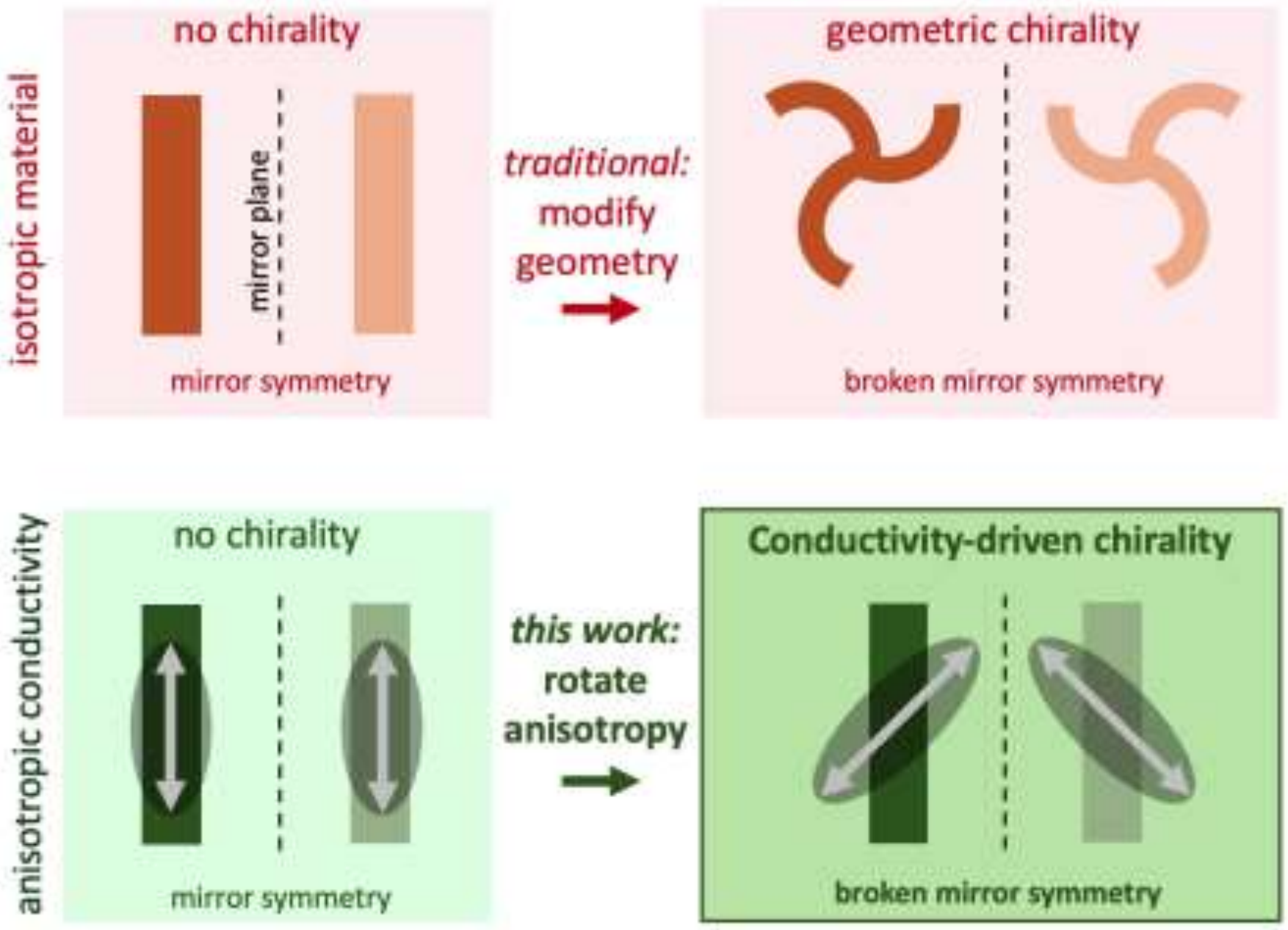

***Figure 1. Concept illustration of conductivity-driven LSPRs chirality.*** *The top row illustrates chirality via geomtry, while the bottom row shows how mirror symmetry can instead be broken in a geometrically achiral structure made from a material with anisotropic conductivity and rotating the conductivity tensor.*

## Results

We validate the concept using the conducting polymer poly(3,4-ethylenedioxythiophene) (PEDOT). PEDOT is known to provide optically metallic properties and support plasmonic resonances in the near-infrared region[16-17]. The carrier density can further be electrochemically tuned[18], which have attracted significant interest for application in reconfigurable metasurfaces[19-20]. In this study, we utilize that the conductivity of PEDOT can be made in-plane anisotropic by strain-induced alignment[17]. In brief, we synthesize PEDOT films by vapor phase polymerization (VPP) on an elastomer substrate and partially align the polymer by uniaxial stretching. The aligned PEDOT film can then be transferred to arbitrary substrates followed by dissolving the elastomer. Due to strong *π–π* conjugation and orbital overlap along the polymer backbone, the charge carrier mobility is significantly higher along the chain direction compared with its transverse directions (Fig. 2a). Such in-plane aligned films therefore exhibit anisotropic conductivity, where the conductivity can be 30% higher along the stretching direction[17]. Figure 1b shows how this anisotropic conductivity translates into anisotropic permittivity, for an aligned PEDOT film doped by sulfuric acid. Owing to higher carrier mobility along the alignment direction, which is the principal conductivity axis, the real permittivity exhibits larger negative values along this direction compared with its cross direction. Likewise, the carrier's effective mass ($m^*$) along this direction is lower due to its reversal relationship with carrier mobility[21-22]. This leads to significantly higher plasma frequency along the alignment direction as

$$\omega_p = \sqrt{\frac{ne^2}{\varepsilon_0 m^*}}, \quad (1)$$

where *n* is the charge carrier density, *e* is the elementary charge and $\varepsilon_0$ is the vacuum permittivity. Consequently, the aligned films exhibit in-plane hyperbolic dispersion in the near-infrared, supporting directional plasmonic resonances along the alignment direction. In the near-infrared to mid-infrared, the film transitions to an in-plane elliptical regime ($\varepsilon_{xx} \neq \varepsilon_{yy}$) with negative permittivity, enabling plasmonic resonances along both axes but at different conditions.

Key for this study is that the in-plane anisotropic properties of the PEDOT can be utilized to induce chiral response in geometrically mirror-symmetric nanostructures. Optical simulations of individual rectangular nanostructures indicate clear optical chirality that can be continually tuned without altering the geometry and only rotating the conductivity tensor relative to the shape symmetry axis (Fig. 2c)[23]. Such rotation smoothly tunes the g-factor (see Methods) calculated from the scattering cross-section of single nanoantenna, from negative values, through zero, to positive values (Fig. 2c, Supplementary Fig. S2). This behavior contrasts sharply with geometry-induced chiral plasmonic systems, in which variation in chiral response requires changes to the nanostructure geometry.

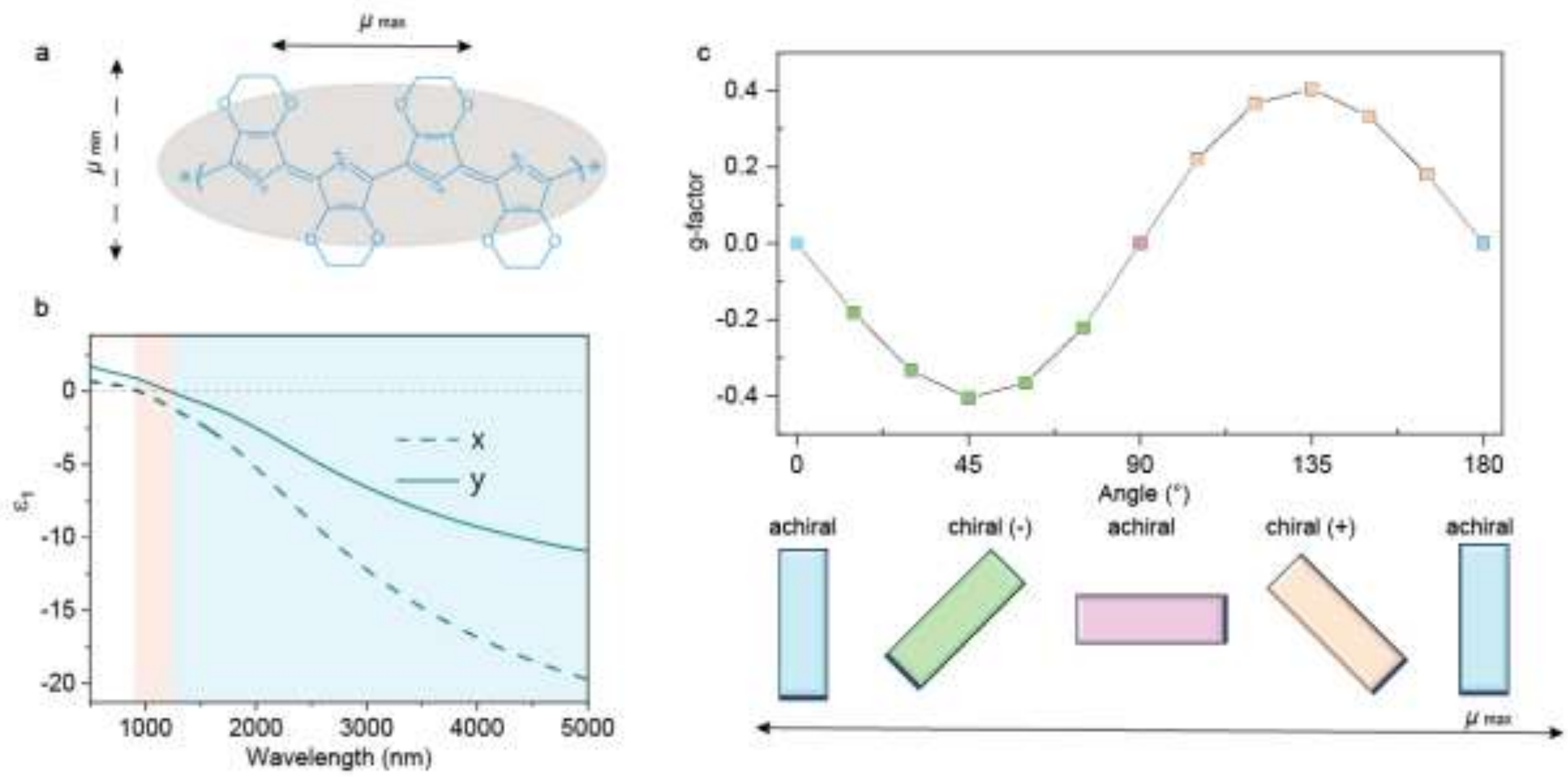

***Figure 2. Planar chirality in plasmonic PEDOT structures with mirror-symmetric shapes.*** ***a)*** *Molecular structure of PEDOT. The shaded elliptical shape illustrates the conductivity anisotropy, where the maximum and minimum carrier mobilities lie along and perpendicular to the polymer chain direction, respectively.* ***b)*** *Real part of the in-plane permittivity of aligned PEDOT polymer in its doped state. || means along the alignment direction which is also the principal conductive axis; ⊥ is normal to the alignment direction. The orange and blue regions indicate the spectral regions that support unidirectional and anisotropic plasmonic resonances, respectively.* ***c)*** *Simulated g-factor (at 1.8 μm) of a single rectangular nanoantenna (200 nm (width) x 600 nm (length) x 150 nm (height)) composed of anisotropic conductive PEDOT, simulated by varying the angle between the principal conductive axis relative to the short side of the rectangle as shown in the scheme.*

We fabricated rectangular nanostructure arrays using electron-beam lithography (EBL) on strained PEDOT films attached on glass substrates. To exclude any chirality originating from geometric or lattice asymmetry, we used identical fabrication conditions to fabricate arrays with different angles between the material's principal conductivity axes and the shape's symmetry axes, achieved by rotating the polymer alignment direction relative to the e-beam writing field. Scanning electron microscope (SEM) images in Fig. 3a confirm that the fabricated patterns are identical, with only varying polymer alignment directions as: 0°, 90°, 45°, and −45° relative to the short side of the rectangles. Polarization-dependent linear extinction spectra show plasmonic resonance peaks and with pronounced differences for the different alignment angles (Fig. 3b, Supplementary Fig. S3). For both 0° and 90° conductivity alignments, the extinction at polarizations $P_{45}$ and $P_{-45}$ are the same, corresponding to no optical chirality. The reason is that the anisotropy in conductivity and shape share the same pair of orthogonal principal directions (they are aligned), such that mirror symmetry is maintained. The behavior is fundamentally different for the samples with conductivity axis ±45° relative to the structural axes (misaligned). In these cases, the conductivity-anisotropy breaks the in-plane mirror symmetry along ±45°, resulting in resonance anisotropy between $P_{45}$ and $P_{-45}$ excitations. CD measurements verify that such breaking of mirror symmetry via conductivity anisotropy can induce optical chirality (Fig. 3c). We first note that 0° and 90° conductivity alignments does not provide any CD signal, confirming that these configurations remain optically achiral. By contrast, the metasurfaces with ±45° conductivity alignment display clear and opposite CD signals, forming an enantiomeric pair despite sharing identical geometry. Optical simulations reproduce these findings (Fig. 3d). The correlation between the wavelengths of maximum CD signal and the plasmonic resonances implies that the observed CD originates from mirror symmetry breaking of the plasmonic resonances. Optical simulations using an ideal anisotropic Drude medium further corroborate that the planar chirality originates from a general symmetry mechanism rather than from any material-specific band structure (Supplementary Drude material, Fig. S11). The Drude nanoantenna reproduces the chiroptical response, confirming that it results solely from the interplay between anisotropic charge transport and the structural symmetry axis. Because similar anisotropy occurs in many other molecular, low-dimensional, and semiconductor systems,[24-26] our approach provides a general route toward ultrathin chiral photonic and optoelectronic systems.

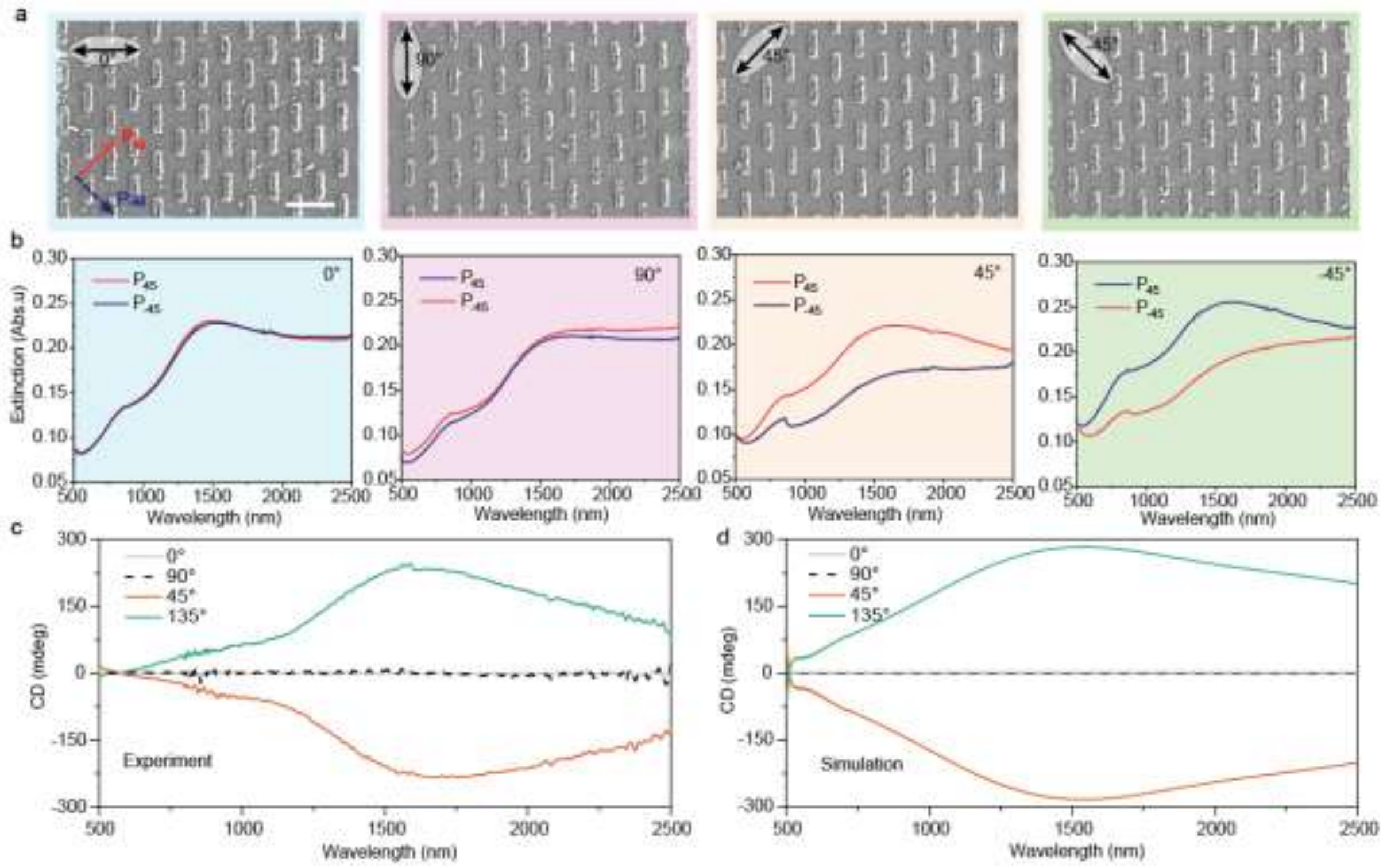

***Figure 3. Different optical polarization states hidden in the same sape. a)*** *SEM images of metasurfaces with rectangles oriented at 0°, 90°, +45° and −45° with respect to the material's principal conductive axis (indicated by the black arrow). The scale bar is 1 µm.* ***b)*** *Extinction spectra at different linear polarizations for the four metasurfaces, with the corresponding polarization directions indicated by the colored arrows in (a).* ***c)*** *and* ***d)*** *Measured (c) and simulated (d) CD spectra of the four metasurfaces. The simulated nanostructures are 150 nm in width, 400 nm in length and 120 nm thick.* ***e) and f)*** *FDTD simulation of spatial mapping of the difference in near-field intensity ($\Delta|E|^2$) between LCP and RCP excitation for the same rectangle nanoantennas made from isotropic (e) and anisorpic (f) conductive PEDOT. Note that the mapping for RCP excitation undergoes a diagonal-mirror operation. The elliptical shapes in (a), (e) and (f) depict the tensor of the in-plane conductivity.*

The role of conductivity anisotropy for the chiral plasmonic response can be further understood from simulated near-fields of nanorectangles made from either isotropic PEDOT or anisotropic PEDOT with the principal conductivity axis rotated by 45° with respect to the geometry axis (Fig. 4; Supplementary Near-field mapping simulations, Fig. S5,S6). For the isotropic PEDOT nanorectangle, the phase delay as induced by linear birefrigence is the same under LCP and RCP excitation, due to the mirror symmetry of the system (Fig. 4b, note here that the RCP phase was shifted by π/2 to match the phase convention of the incident source). Moreover, the optical near field intensity for excitation with LCP and RCP are mirror images of each other (Fig 4c, Supplementary Fig. S5c). The structure therefore provide equal transmission of LCP and RCP light and no optical chirality. The results are fundamentally different for the nanorectangle with anisotropic conductive PEDOT misaligned at 45°, which instead show clear chiroptical effects. First, the near-field phase response under LCP excitation cannot be made identical to that of RCP excitation through mirror operation along any structural symmetry axis (Fig. 4e, Supplementary Near-field mapping simulation and Fig. S6). This asymmetric phase difference between LCP and RCP excitation implies unequal polarization modification for LCP and RCP light, demonstrating circular birefringence effects (CB). Likewise, the optical nearfield intensity under LCP and RCP excitation are not mirror images as for the isotropic PEDOT case, but instead show clearly different hot spot field distribution and enhancement factors as clear from their difference (Fig. 4f, Supplementary Fig. S6c). These effects can explain the CD effect measured in the far field.

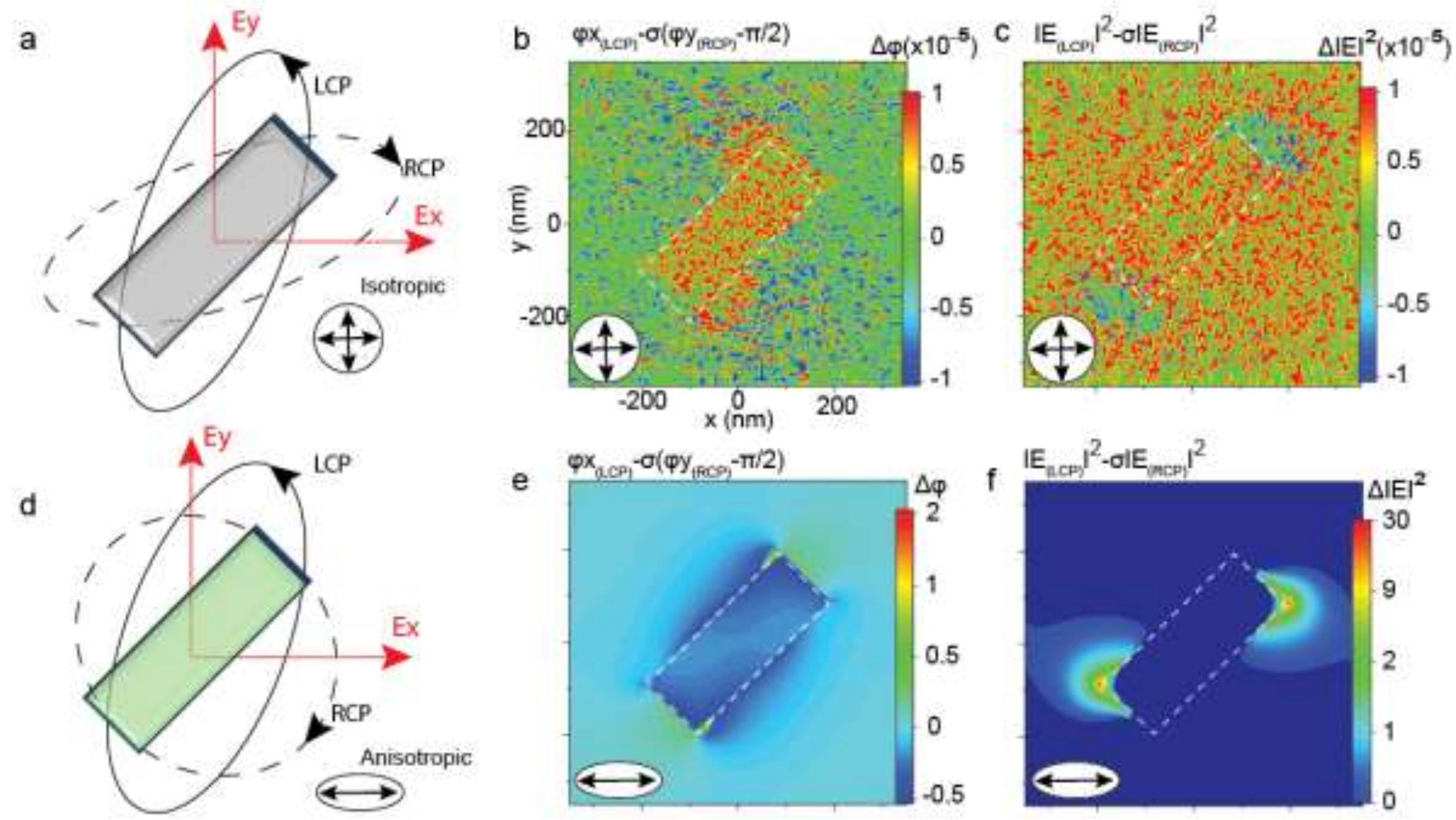

***Figure 4. Near field maping of the CB and CD effects.*** *a,d) Schematic illustrations of polarization conversion in a rectangular nanoantenna composed of either isotropic PEDOT (a) or anisotropic conductive PEDOT (d). The circular and elliptical shapes next to the nanostructures indicate the conductivity profile, with the double arrows marking the principal conductivity axes. b,e) Simulated spatial mapping of the phase difference* $\Delta\varphi$ ( $\Delta\varphi =$

*$\varphi x_{LCP} - \sigma_d(\varphi y_{RCP} - \pi/2)$ between $E_x$ under LCP excitation and $E_y$ under RCP excitation for isotropic PEDOT (b) and anisotropic PEDOT (e). For consistency with the excitation configuration, the $E_y$-field under RCP excitation was mirrored along the diagonal direction ($\sigma_d$) and shifted by $-\pi/2$. c,f) Simulated spatial mapping of the difference in near-field intensity ($\Delta|E|^2$) between LCP and RCP excitation for isotropic PEDOT (c) and anisotropic PEDOT (f). Note that the mapping that under RCP excitation undergoes a diagonal-mirror operation, denoted by $\sigma$. The circular shape in (b) and (c) indicate isotropic conductivity while the elliptical shape in (e) and (f) shows the anisotropic conductivity profiles.*

The principle of conductivity-driven chirality illustrated above is not limited to rectangles, but any planar nanostructure with noncircular geometry (even a square) can exhibit asymmetric interactions with chiral light if their principal conductivity axes are not parallel or normal to any geometric symmetry axis (Supplementary Fig. S1). Following this principle, we designed a grating-type metasurface which achieves much stronger and ultrabroadband chiral optical response. We fabricated two enantiomeric metasurfaces with grating lines oriented either 45° or -45° relative to the polymer's principal conductivity axis (Fig. 5a). The film-like quasi-continuum absorption features along the grating lines, as evidenced by linear transmission spectra (Supplementary Fig. S4), can facilitate substantially broadened CD response compared with the nanorectangles investigated above, and also provide additional benefits such as possibility for electrical contacts and bias without supporting electrodes. Remarkably, a single layer (~120 nm thick) and geometrically symmetric polymer grating metasurface exhibits maximum circular dichroism exceeding 700 mdeg under normal incidence (Fig. 5d). This means that even this structurally simple design of a linear grating can surpass the chiroptical performance of far more complex metal plasmonic architectures[13, 27]. Considering the intrinsically weaker plasmonic resonance of conducting polymers compared with metals, the results further highlight the design advantage offered by exploiting anisotropic charge transport.

To experimentally verify that this broadband CB and CD response extends into the infrared region, we performed full Mueller matrix transmission ellipsometry at normal incidence on metasurfaces prepared on IR transparent calcium fluoride ($CaF_2$) substrates (Supplementary Fig. S7, S9 and S10). Figure 5b shows the dispersion of the M14 and M41 matrix elements for a grating surface with 45° lines relative to the main conductive axis that represents the CD response[28-30], while Figure 5c shows dispersion of the M23 that represents the CB response[28-30], both showing non-zero values in an ultrabroad range from 0.6 to 10 $\mu$m. Moreover, the M14 and M41 elements exhibit a reversed signal, which indicates that the chirality of the metasurface is expected to invert under circularly polarized illumination incident from the opposite direction. Such behavior is indeed confirmed by the reversal of the measured CD signal by illuminating the metasurface from the other side (Fig. 5e). The asymmetric transmission of CPL incident from opposite sides of the surface arises from the metasurface's planar optical asymmetry, while both the structure and conductivity individually remain mirror-symmetric along the surface normal direction. Consequently, an apparent non-reciprocal optical response emerges, similar to the behavior of planar chiral metasurfaces formed by curved lines with explicit symmetry breaking[31-32]. We note that this effect is fully compatible with reciprocity constraints, because no time-symmetry breaking is involved. The Mueller matrix analysis (Supplementary Mueller matrix) also reveals multidirectional linear birefringence (LB) and linear dichroism (LD) in the symmetric structures made of anisotropic PEDOT (Supplementary Fig. S8). For metasurfaces containing grating lines fabricated from isotropic PEDOT, the optical axis follows the grating direction, resulting in LB and LD responses aligned along 0° and 90° polarization axes, which maximized near the plasmonic resonance peak. By contrast, metasurfaces fabricated from anisotropic PEDOT with grating lines 45° oriented with respect to the principal conductivity axis also show clear LB and LD along the ±45° directions because of the conductivity anisotropy. While previous works utilized multilayer-induced effective anisotropy to facilitate LB–LD coupling between adjacent layers[4, 33], our approach achieves the coexistence of LB and LD within a single-layer metasurface arising from geometry and conductivity, respectibley, resulting in asymmetric chiral light transmission and apparent non-reciprocity[32].

An attractive added feature of conducting polymers is that they are electrochromic, with tunable absorption[34] or CD response for polymer systems with chiral geometry[35-37]. Here, we utilize the fact that the permittivity of PEDOT can be switched all the way from negative values (optically metallic) to positive values (dielectric) by reducing its charge-carrier density by de-doping[16, 38]. For our metasurfaces, this switches off their chiral plasmonic resonances and corresponding CD response. Figure 5f demonstrates this dynamic feature for a metasurface exposed to vapor from highly branched PEI, which de-dopes the polymer and leads to a complete switch-off of the CD signal. Without the need for intrinsic chiral molecules or chiral solvents, the method reported here thus represents a fundamentally different and more flexible approach to tunable chirality with conjugated polymers.

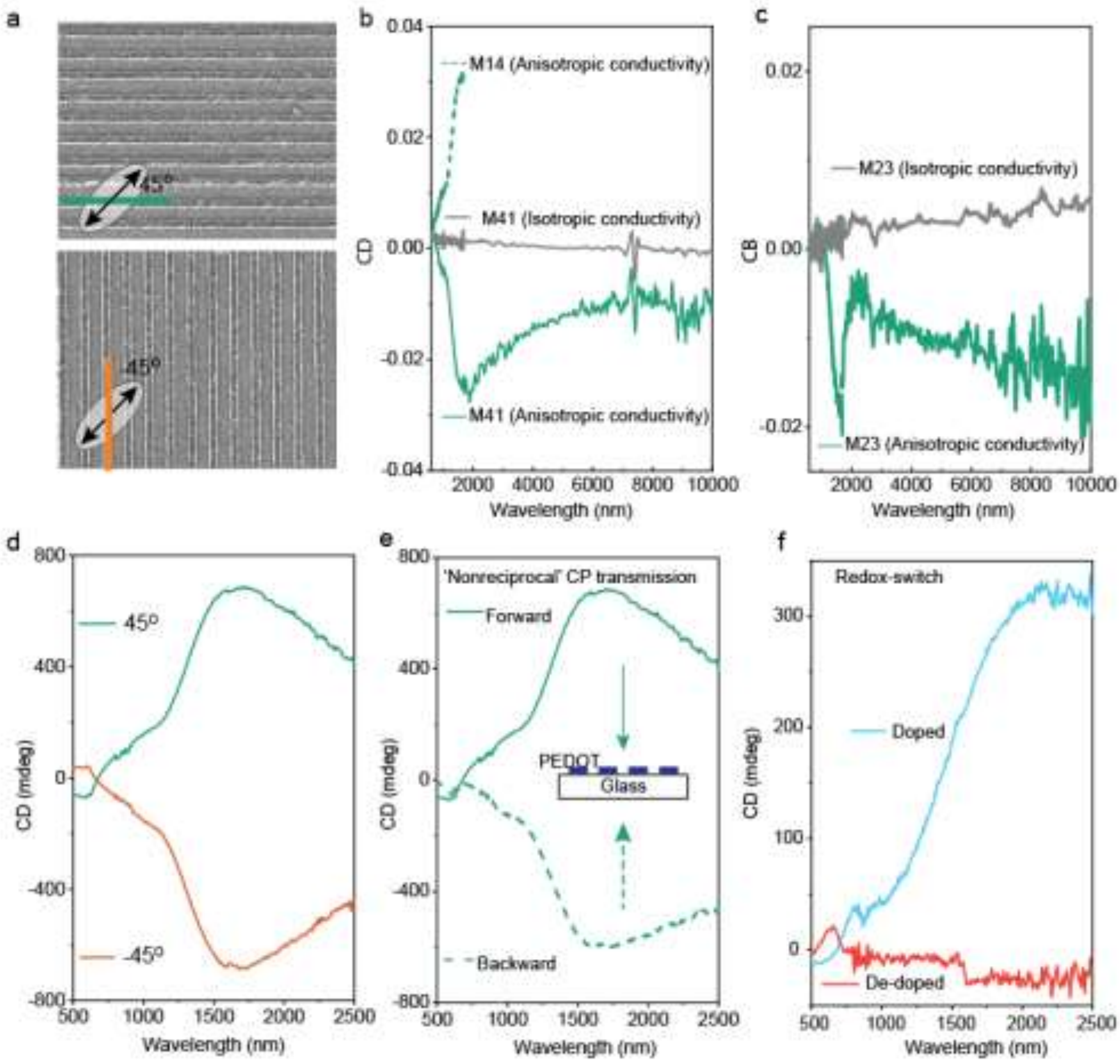

***Figure 5. Tunable and broadband circular dichroism of grating metasurfaces.*** ***a)*** *SEM images of metasurfaces with grating lines oriented at +45° and −45° with respect to the material's principal conductive axis (indicated by the black arrow). The elliptical shape depicts the tensor of the in-plane conductivity. The scale bar is 1 μm.* ***b)*** *The* $M_{14}$ *and* $M_{41}$ *elements of the Mueller matrix for a grating metasurface with lines oriented at 45° with respect to the material's principal conductive axis. The grey line shows results for a grating metasurface with the same pattern but made from isotropic PEDOT.* ***c)*** *The* $M_{23}$ *elements of the Mueller matrix for a grating metasurface with lines oriented at 45° with respect to the material's principal conductive axis. The grey line shows results for a grating metasurface with the same pattern but made from isotropic PEDOT.* ***d)*** *Corresponding CD spectra of the metasurfaces shown in (a).* ***e)*** *CD spectra of the same metasurface measured under forward and backward light propagation. The slight difference in intensity is caused by the substrate.* ***f)*** *CD spectra of a metasurface measured in the oxidized (doped) and reduced (de-doped) states.*

Finally, we will demonstrate that the presented concept opens new opportunities for phase-gradient metasurfaces. For isotropic conductors, including PEDOT, it is well known that the rotation of rectangular nanobars can generate a geometric (PB) phase. This local phase control has been widely employed in metasurfaces to provide functions including beam deflection, lensing or holography.[19] In a conventional PB-phase design (with isotropic conductor), the phase varies linearly from 0 to $2\pi$ as the rotation angle increases from $0°$to $180°$. Meanwhile, the cross-polarization conversion efficiency (LCP to RCP and RCP to LCP, respectively) is chirally symmetric and independent of rotation angle. Figure 6a illustrates these features for nanobars made from isotropic PEDOT.

Nanobars made from anisotropic PEDOT shows a fundamentally different behavior. They also provide 0 to $2\pi$ phase control through the rotation angle, but in addition they enable chiral symmetry-breaking manifested in a polarization conversion efficiency that is both spin and angle-dependent (Fig. 6b). For example, the LCP to RCP conversion efficiency shows a clear maximum at 45° rotation angle and minimum at 135° rotation angle, while the RCP to LCP conversion efficiency shows the opposite behavior. At 0° and 180°, the conversion efficiency is instead achiral. To understand these dependencies, we recall from Fig. 2c that the g-factor of each rectangular meta-atom follows approximately $g_{(\theta)} \propto -\sin(2\theta)$, where θ denotes the rotation angle of the meta-atom with respect to the (macroscopically) fixed principal conductivity axis. The local chirality is thereby intrinsically linked to the geometric phase. Since the PB phase ($\phi_{(x,y)}$) encoded by each meta-atom is given by $2\theta$, the corresponding chirality distribution is encoded in the phase profile *via* $g_{(x,y)} \propto -\sin[\phi_{(x,y)}]$.Hence, any PB metasurface made from these metaatoms inherently carries a deterministic spatial chirality distribution accompanying the prescribed phase profile.

To illustrate capabilities opened through this added spin-dependency, we consider beam deflection metasurfaces based on a linear π phase shift (Figure 6c-6e). When using isotropic PEDOT for such metasurfaces, the far-field diffraction patterns are independent on the initial phase of the first meta-atom in the super cell. Moreover, the diffraction patterns generated under LCP and RCP excitation exhibit mirror symmetry. By contrast, metasurfaces based on PEDOT providing anisotropic conductivity, the diffraction response is governed not only by the geometric phase gradient but also by the spatial distribution of local chirality across the metasurface. This originates from the different dependences of these two parameters on the meta-atom rotation angle: the PB phase gradient exhibits a linear dependence, whereas the chirality gradient follows a sinusoidal dependence (Figure 6b). Consequently, even a super period within the same gradient and phase range, the diffraction behavior can be strongly modulated by the initial rotation angle of the first meta-atom in the super period unit cell. Furthermore, for an identical phase-gradient range, the diffraction patterns under LCP and RCP excitation can be engineered to either exhibit mirror symmetry (Figure 6d) or become clearly distinct from each other (Figure 6c,e), depending on the selected chirality-gradient distribution. These results reveal that the rotation-angle-induced chirality gradient, intrinsically linked with the PB phase, provides an additional degree of freedom for geometric metasurface design, enabling new opportunities for multifunctional PB metasurfaces beyond conventional phase-gradient engineering.

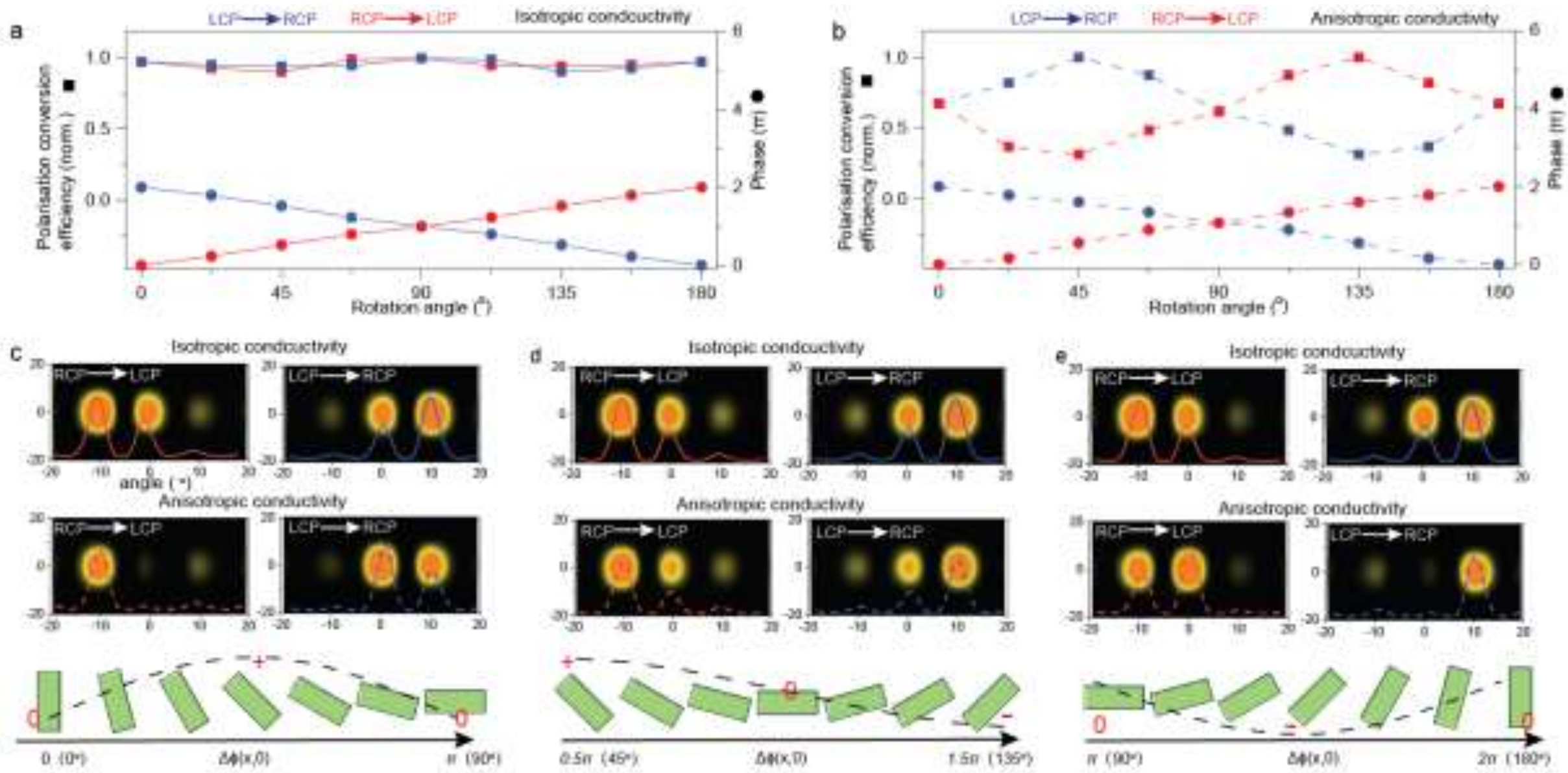

***Figure 6. Chirality-gradient geometry metasurfaces based on PB phase. a,b,*** *Simulated cross-circular polarization conversion efficiency and geometric phase as a function of the nanoantenna rotation angle for isotropically conductive (a) and anisotropically conductive (b) PEDOT nanoantennas.* ***c–e,*** *Simulated far-field diffraction patterns of beam-steering gratings with the same overall π geometric-phase gradient but different PB phase ranges: 0– π (c), 0.5π–1.5π (d), and π–2π (e). The blue and red curves indicate the relative diffraction intensities of the diffraction orders under opposite circularly polarized excitations. The dashed lines in the schematic illustrations indicate the variation in chirality, accompanied by the 0, + and - signs. The principal conductivity axis was set to the x-direction. The size of the nanorectangles is 450nm in length, 150nm in width and 150nm in height, with 1μm centre-to-centre distance 7 elements in each super period. The illumination was set to normal incidence at 1.25 μm wavelength.*

## Discussion

This study demonstrates that conductivity anisotropy provides a new degree of freedom for realizing continuously tunable spin-selective plasmonic resonances in geometrically invariant meta-atoms. The underlying symmetry principle is general and should be applicable to a wide range of anisotropic conductive materials. This generality arises because the chiroptical response originates from the coupling between linear birefringence and linear dichroism, which are governed by the geometry and the anisotropic conductivity, respectively, rather than by material-specific electronic properties. This mechanism enables nanoscale chirality in single symmetric meta-atoms, eliminating the need for chiral geometry or large-area asymmetric meta-atoms arrangements, thereby opening a new conceptual route toward chiral metasurfaces that to not rely on geometric asymmetry. Beyond conductivity-induced chirality, our results reveal a previously unexplored connection between geometric phase and optical chirality. In conventional PB metasurfaces, the rotation angle of each meta-atom encodes only the local geometric phase. Here, because the local chirality is determined by the same rotation angle relative to the principal conductivity axis, every meta-atom simultaneously encodes both the geometric phase and its local chiroptical response. Consequently, a prescribed phase profile is intrinsically accompanied by a deterministic spatial chirality distribution, introducing an additional design parameter that is absent in conventional metasurfaces based on metaatoms of isotropic materials. This coupling enables simultaneous wavefront shaping and spin-selective light manipulation through a single rotational parameter and provides new opportunities for multifunctional geometric-phase devices. In summary, by coupling geometric coordinates with material coordinates, it establishes a new route toward programmable chiral nanophotonics and extends geometric-phase metasurfaces from phase-only design to simultaneous phase–chirality engineering.

## Materials and Methods

***Metasurface fabrication:*** *Styrene ethylene butylene styrene (SEBS) was used as a soft substrate to induce PEDOT alignment through mechanical stretching. A 50µm thick SEBS layer was spin-coated onto a glass substrate and dried at 75 °C. The thin film PEDOT was then grown on SEBS by vapor-phase polymerization of its monomer 3,4-ethylenedioxythiophene with iron (III) trifluoromethane sulfonate ($Fe(OTf)_3$) as oxidant. The synthesized PEDOT thin film was further doped by immersion in 1 M sulfuric acid. The vapor phase polymerization process was then repeated twice to obtain a total film thickness of approximately 60 nm. Subsequently, the SEBS support layer was peeled off from the original glass substrate. The SEBS film was then uniaxially stretched by 25%, aligning the PEDOT chains in-plane along the stretching direction. The stretched film was inverted and attached onto another glass substrate with the PEDOT side facing the glass. Finally, the SEBS layer was dissolved in toluene, leaving a thin, aligned PEDOT film attached to the glass surface.*

*EBL was employed to fabricate the PEDOT metasurfaces. A ZEP 520A resist (Zeon Europe GmbH) was spin-coated onto the PEDOT films at 1500 rpm for 1 min and baked at 140 °C. A thin layer of conductive polymer was spin coated on the top of ZEP resist to improve the EBL quality. EBL exposure was carried out at 50 kV with a dose of 100 µC $cm^{-2}$. The substrate was rotated relative to the write field so that the patterned structures attained the desired angle with respect to the main conductive direction of the material. After e-beam exposure, the conductive polymer layer was removed by dry etching with $O_2$ and then was developed in ZED-N50 for 1 min. Reactive ion etching (RIE, Advanced Vacuum) was then performed for 140 s using 50 W RF power, 200 mTorr chamber pressure, and an $O_2$ flow rate of 100 sccm. Finally, the residual resist on top of the PEDOT nanostructures was removed by rinsing with AR 300-72 (NEP, Allresist GmbH). The fabricated metasurfaces are 6x6 mm.*

***Ellipsometry measurements:*** *The metasurfaces were fabricated on $CaF_2$ substrates for spectroscopic ellipsometry measurements. The Mueller matrices were acquired using spectroscopic ellipsometry. All measurements were performed at normal incidence to avoid any extrinsic symmetry breaking. For the wavelength range of 0.4–1.69 µm, UV–vis–NIR data were collected using a dual-rotating compensator ellipsometer (RC2, J.A. Woollam Co., Inc., Lincoln, NE, USA). For the 1.69–10 µm range, infrared measurements were performed using a J.A. Woollam Co. IR-VASE spectroscopic ellipsometer. The Mueller matrix was extracted and analyzed using the CompleteEASE software (J.A. Woollam Co.) based on the acquired ellipsometry data.*

*The Mueller matrix (M) describes how a sample transforms the Stokes vector of the incident light:*[28]

$$S_{out} = MS_{in}$$

*Here, each element $M_{ij}$ tells us how one polarization component is converted into another.*

*The fabriacted grating metasurface is effectively uniform along the surface-normal direction, meaning that its optical response does not involve multiple scattering or layer-to-layer random phase mixing. It therefore behaves as an anisotropic but non-depolarizing thin film. In this case, it is appropriate to analyze the differential Mueller matrix, defined as* [29-30]

$$dM = log(M)$$

*Here, dM represents the infinitesimal polarization changes introduced by the meatasurface, and each optical effect (LD, LB, CD, CB) corresponds to a specific element:*

$$dM = \begin{bmatrix} A & -LD & -LD' & CD \\ -LD & A & CB & LB' \\ -LD' & -CB & A & -LB \\ CD & -LB' & LB & A \end{bmatrix}$$

*Where CD and CB are the circular dichroism and circular birefringencem; LD and LD' are linear dichroism along 0°/90° and ±45°; LB and LB' are linear birefringence along 0°/90° and ±45°; A is absorbance.*

***Linear Spectroscopy:*** *The linear polarization dependent spectra were obtained from UV-Vis spectrometer (Lambda 900 and Lambda 1050+ from Perkin Elmer) and Fourier Transform Infrared Spectroscopy (Perkin Elmer Spectron 3) at the normal transmission state, both equipped with linear polarizers. The sample was rotated to get the spectra at a specific polarization.*

***CD spectroscopy:*** *CD spectra were characterized using a high-performance circular dichroism spectrometer (JASCO J-1700). Measurements were conducted over a wavelength range of 2500–300 nm with a data pitch of 5 nm, a scanning speed of 500 nm min$^{-1}$, a bandwidth of 1 nm, and a data integration time of 1 second. The samples were mounted on a diaphragm, with the beams incident normal to the samples. Baseline correction was carried out using the same blank glass substrate before data acquisition. Each spectrum was collected three times, and the averaged result was used to ensure high signal accuracy.*

***FDTD simulation:*** *The finite-difference time-domain (FDTD) method was used for numerical simulations, using the commercial software Ansys Lumerical FDTD. In the simulations, circularly polarized light is implemented using two light sources that are settled at the same position with polarization along the x- and y-directions, respectively.*

*For LCP, the phase of the x-polarized component is set to 0°, and the y-polarized component is −90°. For RCP, the x-polarized component is set to 0°, while the y-polarized is +90°. To evaluate the scattering cross-section of the isolated nanostructure, we employed a Total-Field Scattered-Field (TFSF) source. A box monitor was placed outside the TFSF boundary to capture only the scattered field components. The g-factor of the nanostructure was calculated from*

$$g = 2 * (S_{LCP} - S_{RCP})/(S_{LCP} + S_{RCP}),$$

*where $S_{LCP}$ and $S_{RCP}$ are the scattering cross-section under LCP and RCP excitation. To evaluate the CD of the metasurfaces, the simulation was carried out with periodic structure on a substrate with index fixed at 1.5. Plane wave excitation was employed in combination with periodic boundary conditions in the FDTD simulation. The excitation light propagated in the -z direction, and a plane monitor was placed after the sample to measure the transmitted spectra (T), ensuring consistency with standard CD spectrometer measurements. The CD (mdeg) value was calculated by*

$$CD = 32980 * log_{10}(T_{RCP}/T_{LCP}).$$

*The period is the same along both the x- and y-directions, so that the lattice remains fully symmetric and does not contribute to any symmetrical breaking. Any observed chiral response therefore originates solely from the anisotropy of single structures rather than the lattice.*

### *Acknowledgements*

*The authors acknowledge support from the European Research Council (Consolidator Grant, 101086683), the Swedish Research Council (VR, Consolidator Grant, 2020-00287), the Knut and Alice Wallenberg Foundation (Wallenberg Academy Fellow, 2019. 0163), and the Swedish Government Strategic Research Area in Materials Science on Functional Materials at Linköping University (Faculty Grant SFO-Mat-LiU No. 2009 00971).*

### *Contributions*

*Y. D. and M. J. conceived the study; Y. D. designed the experiments, performed the FDTD simulations with H.M.S and theoretical analysis, fabricated the metasurfaces by EBL with assistance from D. L., and wrote the initial manuscript together with M. J.; L. L. carried out the CD measurements with assistance from Z. Z and W. W.; S. K. prepared the aligned PEDOT films and performed the linear spectroscopy and ellipsometry measurements; R. M. analyzed the Mueller matrices. X.X. developed the recipe and prepared the top conductive resist (E-HERO) to enhance the nanostructure quality of EBL; Y. D. and M. J. discussed the project.*

### *Data availability*

*The data presented in this study is provided either in the Supplementary Information or at Zenodo at XXX (link provided upon acceptance).*

### Conflict of interest

*The authors declare no competing interests.*